\documentclass[linenumbers]{aastex631}

\usepackage{booktabs}
\usepackage{tabularx}
\usepackage{ragged2e}
\usepackage{array}        

\begin{document}

\title{Was there a (super)nova in 1408?}

\author[0009-0003-0458-5210]{Boshun Yang}
\affiliation{University of Science and Technology of China \\
No. 96 Jinzhai Road \\
Hefei, 230026, China}

\author{Nikolaus Vogt}
\affiliation{Universidad de Valparaíso \\
Instituto de Física y Astronomía\\
Av. Gran Bretaña 1111\\
 Valparaiso, Chile}

\author[0000-0001-5384-7545]{Susanne M Hoffmann}\thanks{Corresponding author: susanne.hoffmann@uni-jena.de}
\affiliation{University of Science and Technology of China \\
No. 96 Jinzhai Road \\
Hefei, 230026, China}\affiliation{Michael-Stifel-Center Jena \\
 Friedrich Schiller University Jena, JenTower,\\
 07737 Jena, Germany}

\collaboration{20}{(AAS Journals Data Editors)}

\begin{abstract}
The 1408 CE “guest star" recorded in Chinese historical texts presents a compelling case for identifying a historical stellar transient. While previous studies debated its nature as a meteor, comet, or nova, we re-evaluate the event using original Ming Dynasty records, including a newly found memorial from the imperial court. The object, described as stationary for over ten days, yellow, and luminous (resembling a “Zhou Bo virtue star"), is inconsistent with cometary behavior. Positional analysis locates it near the \textit{Niandao} asterism (modern Cygnus-Vulpecula region) within the Milky Way, with a derived brightness of $-4$ to 0~mag. Light-curve stability over ten days and color descriptions align with a slow nova or a supernova. We cross-correlated the historical coordinates with modern catalogs and found a few possible counterparts. Among them, CK Vul -- a luminous red nova remnant from 1670 to 1672 -- is the most interesting candidate. Could its progenitor system have experienced a precursor classical nova eruption circa 1408 prior to the merger $\sim200$ years later? We also examine cataclysmic variables and planetary nebulae within the 100 square-degree search field, though most lack sufficient brightness or age characteristics. This study emphasizes the value of integrating detailed historical records with contemporary astrophysical data to resolve long-standing controversies over ancient transients. The 1408 event likely represents a rare, well-documented nova, offering insights into pre-modern stellar phenomena and their modern counterparts.  
\end{abstract}


\section{Introduction} \label{sec:intro}
Within the past $\sim90$ years, several supernovae have been successfully discovered among Chinese observations of “guest stars”. \citet{Mayall1942} suggested the identification of the Crab Nebula (M1) with a guest star in 1054~CE, Tycho’s and Kepler’s Supernovae (1572 and 1604) where matched with their counterpart remnants, and the guest star 1181 observed in Korea, China and Japan was recently interpreted as a supernova of a rather special type \citep{ritter2021ApJ}. Historical novae, however, are much more controversial, as the well-known examples of identifications of stars and nebulae with guest stars in 101~CE \citep{Hertzog1986}, 438~CE \citep{Miszalski2016}, 77~BCE \citep{Johansson2007} turned out as probable comets. 

A very interesting case of decade-long discussion is given by the guest star 1181 observed in China, Korea and Japan. \citet{Stephenson1976} and his subsequent work have suggested the pulsar wind nebula PWN 3C\,58 for the identification. However, \citet{Bietenholz2006,Bietenholz2013} have argued that (a) the nebula is too old, and (b) in 1181 the pulsar would not have been in the center of the nebula. Despite these strong arguments, scholars kept using Stephenson’s identification \citep{Kothes2010, Kothes2013} until \citet{ritter2021ApJ} suggested a new identification, and it turned out to be a rather special type of supernova. 

A similarly controversial and yet undecided debate has been carried out concerning a guest star in the 15th century. The debate is summarized by \citet{Wang2016}: \citet{Li1978} had proposed the case by connecting records of guest stars in September and October 1408 and identifying them as one and the same object in the sky. \citet{Imaeda1980} additionally provided two brief records from Japan in July of the same year regarding the "guest star," arguing that they referred to the same guest star as the ones recorded in September and October in China. \citet{Stephenson1986} provided a detailed argument that the guest stars observed in September and October were actually meteors or comets, and that the Japanese records were unrelated to the two Chinese accounts \citep{Stephenson2005}. \citet{Wang2016} examined the evidence presented by both parties, endorsing Stephenson’s conclusions regarding the Japanese records and the nature of the September guest star in the Chinese records. However, they insist that the possibility of the October guest star being a nova or supernova could not be excluded, although there were several meteors and comets reported in the same year.

In this contribution, we discuss the remaining record from October 1408, that cannot be refuted, as a nova candidate. \citet{Stephenson1986} argued that the absence of a given duration may suggest that it was visible for only a single night. In such a short time frame, a comet could also appear stationary, but comets are typically visible for more than one night. Hence, we looked for more observational records of this celestial guest. 

\section{Analysis of the phenomenon reported in October 1408 (China).} \label{sec:mainpart}
\subsection{Original Records } \label{subsec:Original Records}
Up to now, only the short version of the record has been used. For instance, in \citet[142]{Xu2000}, it reads: 

\begin{itemize}
 \item[] \indent “Emperor Chengzu of Ming, 6th year of the Yongle reign period, 10th month, day \textit{gengchen }[17]. In the night, at the zenith, southeast of \textit{Niandao}, there was a star like an oil-cup of a lamp. It was yellow and shiny bright. It emerged but did not move. It was said to probably be a ZHOU BO, a star of virtue." 
    
\raggedleft{[\textit{Ming Taizong shilu}] ch. 84; [Guo que] ch. 14. }
\end{itemize}

The translation “shiny bright” had alternatives in \citet{Wang2016}, but the explicit statement that the object did not move may point to any sort of nova. The previous entry in \citet{Xu2000}’s list of “guest stars” is a similar record dating to 1404. It reads, “There was a star like a shallow cup southeast of \textit{Niandao}. It was yellow and shiny bright, but did not move” and is quoted from another source [\textit{Ming shi}, \textit{Tianwen zhi}], ch. 27. Although the wording is the same, another date (1404 instead of 1408) is given. \citet{Xu2000} seem to suggest that there were two “shiny bright” guest stars within four years. Yet, the record given in the 2nd year (instead of the 6th) of this reign period could be a writing error by the historical scribe. Thus, there are several problems with these two quotations from \citet{Xu2000}: (1) the date, (2) the translation, and (3) the source; we first discuss these historical questions before deriving the astrophysical consequences. 

\subsubsection{The Sources}
The \textit{Ming shi} is a chronicle written after the end of the dynasty and  published in 1739. It primarily draws its historical materials from the \textit{Ming Taizong shilu} (Veritable Records of Emperor Taizong of the Ming), completed in 1430 and compiled based on original archives from the Yongle reign. Consequently, the \textit{Ming Taizong shilu} (that gives the date 1408) preserved more authentic and accurate information than the \textit{Ming shi}. The \textit{Ming shi}, in turn, was subject to editorial revisions by court historians, leading to the loss of significant information from the original records. However, this had even happened to the \textit{Ming Taizong shilu}. Its substantial editorial revisions resulted in the loss of the original observational data. 

This confusion can be addressed by examining the collected works of individuals who might have personally witnessed the event during the Yongle reign. We found more detailed reports from the Qintianjian (Imperial Astronomical Bureau) included in an original archive: Hu Guang (1370--1418) was one of the contemporary witnesses of the event. Serving as a Hanlin Academy scholar, he was responsible for composing congratulatory memorials and laudatory poems addressed to the emperor following the occurrence of auspicious phenomena, which were often attributed to the emperor’s personal virtue. After the Astronomical Bureau observed the "Zhou Bo Virtue Star," Hu Guang presented a congratulatory memorial to the emperor on behalf of the court officials. To provide readers with a more comprehensive understanding of this star, as well as the circumstances and thoughts of its observers, we translate the content in its entirety rather than extracting some "key" information because this has frequently been part of the interpretation problem in astrophysics. Still, we highlight the passages  that are more relevant to our discussion\citep[623]{huguang1997}: 

\begin{itemize}
  \item[] \indent \textit{Memorial of Congratulations on the Auspicious Star}
  
  \indent  The Qintianjian reported:\textbf{ In the sixth year of the Yongle era, on the sixth day (1408.10.24) of the tenth month at early dusk, a star was observed in the southern region of the \textit{Niandao} in the middle of the sky, appearing as large as a \textit{Zhan} (cup), with pure yellow color, smooth and bright. The star remained stationary and calm over ten days of measurement and observation.} According to the divination texts, this is an auspicious star, known also as the Virtue Star, whose appearance signifies peace throughout the realm. It is said that such a Virtue Star manifests when a ruler governs with propriety, harmonizes rituals and music, and ensures internal and external order while himself enjoying longevity and enduring health. The star was located within the \textbf{celestial division of the lodge Dou (8th lodge)}, precisely corresponding to the imperial domain (according to “field allocation” astrology.) 

    \indent   We, your ministers, have encountered this auspicious sign and respectfully offer our congratulations. \textbf{We humbly thought: the heavenly vault reveals its blessings, with the Virtue Star's glittering brilliance in the middle of the sky; the Silver River (Milky Way) unfolds its splendour, its luminous beauty interwoven along the \textit{Niandao}.} This splendid omen is truly a sign of an enlightened era. We respectfully recognize His Majesty, the Emperor, possesses virtues that reach the heavens and divine accomplishments that surpass all under heaven. With benevolence akin to the heavens and the earth, His Majesty nurtures all beings, bringing perpetual spring; with wisdom that illuminates the farthest corners, His Majesty’s light extends across the four directions. His magnificent governance has refined rituals and music to their height, and his teaching and transformation have brought prosperity and blessings to all. Distant peoples hasten to pay tribute, harvests are abundant, and the years are fruitful. Thus, auspicious signs appear in abundance, and the Virtue Star shines brightly. This is, indeed, the result of harmony among the people below and the response of the heavenly way above. Witnessing such an exalted sign, we, your ministers, cannot contain our joy. To encounter such peace and prosperity as in ancient times is our great fortune, and we celebrate this era of unparalleled governance. Beholding Your Majesty’s countenance at such close distance, we humbly offer wishes for your boundless longevity.
    
    \raggedleft{[\textit{Hu wenmugong quanji}] ch. 9.}
    
    \end{itemize}

This memorial includes (1) a more comprehensive and detailed report from the Astronomical Bureau and (2) a laudatory ode that proves authenticity. The record is written only weeks or months after the observation, probably by an eyewitness. While only the first paragraph has the original observation, the rest of the text emphasizes the great impression it had on the observers. 

According to \citet{chu2022}, during this period, Liu Zhe, the Deputy Director of the Imperial Astronomical Bureau, was tasked with compiling a large volume of comprehensive astrological books. Thus, celestial omens are assumed to have been highly important; we can consider the information reliable.

\subsubsection{(Our) Interpretation}
This memorial addressed to the emperor significantly reinforces the authenticity of this guest star record, effectively eliminating the possibility of fabricated astronomical records often found in official histories. 

Instead of “shiny bright” in \citet{Xu2000}’s translation, we stick to the original Chinese term "guang-run", where "run" refers to a smooth, polished, and lustrous state. The term is typically used to describe white jade and it is associated with a gentle quality and without sharp edges. It does not exactly mean “shiny”, but it is clearly distinct from stars described as having "points" or "spikes" (máng jiǎo). Rays and edges are often associated with dangerous or ominous phenomena.

Note that the supernova of 1006 was recorded with "rays" in Europe, where it was close to the horizon. In contrast, the Imperial Astronomical Bureau of China avoided such a description when they identified the guest star as the Zhou Bo star. Instead, they called it "yellow in color, lustrous" (run-ze), because yellow is a positive color and lustrous does not imply any negative omens. This exemplifies the choice of wording with regard to the divination and the narrative in the chronicle rather than accurate descriptions in the standardized language of astronomical observers. 

\subsection{What astrophysics can we derive?} \label{subsec:astrophysics}
The object is reported stationary for (at least) ten days. Thus, the hypothesis of a comet can be excluded; a stellar transient is highly likely. Also, the object is reported as “calm” for over ten days, which probably means that it did not flicker and they did not observe a significant dimming or change of color. With regard to typical nova lightcurves \citep{Strope2010a, kato2023}, this rarely happens, and it is characteristic for specific types of progenitor stars: plateaus were observed at T Pyx-type recurrent novae and DQ Her-type novae, some of them even with very long plateaus in the order of years. 

\subsubsection{Color}
Color terms are not highly reliable in historical texts. First, because color terms, in general (until today), are not well defined in colloquial language, color perception strongly depends on the environment. Second, there were fewer terms for color in all historical languages (cf. \cite{maclaury2007}), e.\,g. many languages had only one term for green and blue together, and the shades of orange between yellow and red were typically unnamed. Third, the color yellow was auspicious in ancient China, implying that its mention in the omen stresses the positive meaning of the apparition. Again, the 1006 supernova was recorded as whitish-blue in Japan, but in China, it was described as yellow in color to stress its connection to a flourishing period in Chinese history (for both, see \citet[137]{Xu2000}). 

So, the phenomenon of 1408 may have had a bright color, but pure white or bluish would have been interpreted as a bad omen. A reddening due to extinction can be excluded, as the constellation Niandou was “in the middle of the sky”. A large height, close to the zenith, can be confirmed with Stellarium. However, a humid or dusty atmosphere due to common weather conditions may additionally blur and colorize the star. From the description, we only know that the color was not reddish, everything else is possible.  

Although the bright (2~mag) Nova Persei~1901 was described as deeply red by naked eye observers \citep{Archenhold1901b}, most of the novae and the supernovae seem to match this finding. Recent surveys \citep{Craig2025} reveal that the optical color of novae seems to be rather white $(B-V)_0=-0.2 \pm0.3$ at peak and slightly reddens (by $0.2$) during $t_2$, the decline by 2~mag. Thus, they appear bluish-white or white during the initial observation.

\subsubsection{Peak Brightness}
The ode also contains essential information. In addition to the Qintianjian report, Hu Guang claimed that many of the court officials had observed the star, which suggests that the literati at the time might have discussed the star. Hu Guang used what he knew to craft a beautiful portrayal of the night sky. In the memorial, Hu Guang mentioned not only the guest star and the \textit{Niandao} but also the Milky Way, a detail that was not included in the Qintianjian report: astronomers know that the constellation \textit{Niandao} is in the Milky Way, but for an untrained court official, this might stress the beauty of the observation. This suggests that Hu Guang, or individuals within his circle, had really observed the star. They saw the star with the unaided eye in front of the bright background of the Milky Way in the Cygnus-Vulpecula area. Together with the mentioned color, this information points us to first brightness estimate: The object must have had at least the brightness of the modern pole star, as colors are only visible for stars brighter than 2~mag. Furthermore, the term “Zhou Bo star” that they use in the divination was only used for extraordinarily bright phenomena, such as the supernova in 1006 at an estimated apparent magnitude of $-9.5$~mag \citep[132--135]{clark1977}. The description as a Zhou Bo Star of “glittering brilliance”, therefore, suggests a phenomenon as bright as planets (roughly: Venus $-4$~mag, Jupiter $-3$~mag, Mars and Mercury $-2$~mag). 

Even more information can be deduced as ancient Chinese astronomers commonly used objects from daily life to describe the brightness of unusual celestial bodies. The term \textit{zhan} was used in this record, a word frequently employed in the Ming dynasty astronomical records. \citet[47--75]{Wang2008} provided numerous convincing examples to demonstrate that observers assumed a radius of 13 meters for the celestial sphere when they used length rather than angular units for celestial measurements. While the angular size of a (super)nova itself is small (a dot for the eye), atmospheric conditions and the optical limitations of the human eye may lead to an "apparent enlargement effect", namely a cognition of a small disc. The brighter the star, the larger its apparent size. Therefore, descriptions of the size of supernovae, meteors, and other similar phenomena essentially translate to descriptions of their brightness.

The term \textit{zhan} has been overinterpreted as "lamp" \citep{Li1978}, but in ancient China it referred to a cup used for tea or wine or a vessel for lamp oil. \citet[116]{Wang2008} provided the \textit{zhan} diameter of $6-10$~centimeters, corresponding to an angular diameter of $0\fdg26-0\fdg42$ at a distance of 13~meters. The corresponding brightness would be $-5$ to $-7$ mag. However, traditional Chinese tea or wine cups generally have a larger top and a narrower bottom. The diameter he described refers to the upper opening. In contrast, the actual diameter of the cup's base usually ranged from 2 to 4~centimetres, corresponding to an angular diameter of $0\fdg09 - 0\fdg17$ and a brightness of magnitude 0 to $-3$~mag. Therefore, judging from the term \textit{zhan}, the brightness of the 1408 (super)nova should have been between 0 and $-7$~mag.

\subsubsection{Lightcurve}
\begin{figure*}[h]
	{\centering
	\includegraphics[width=0.85\linewidth]{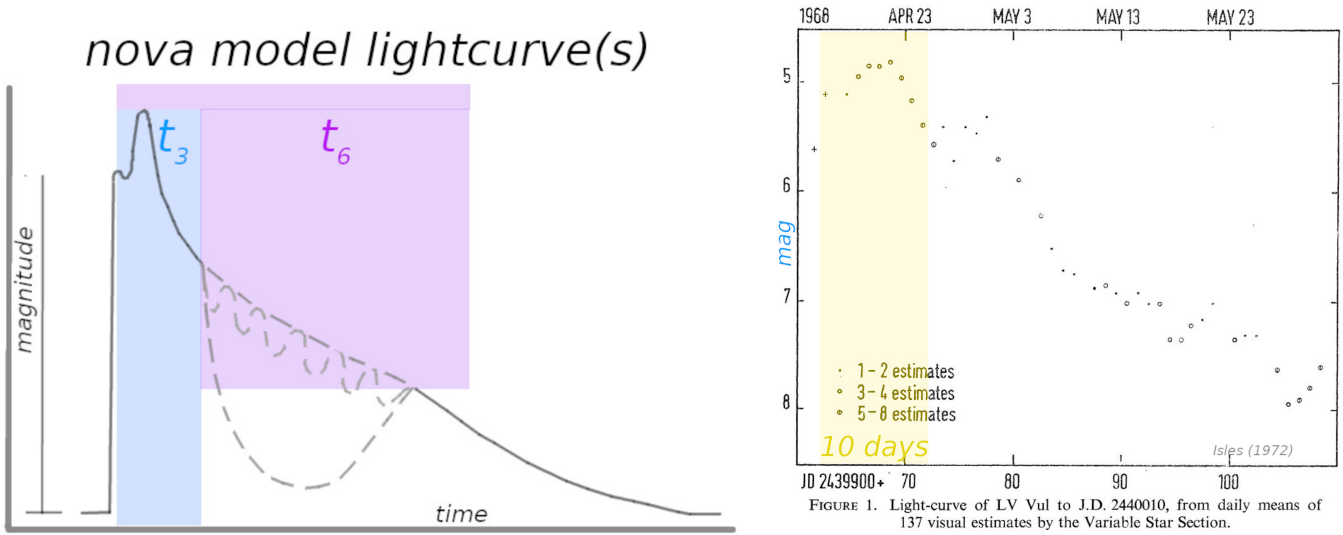}\\}
	\caption{Model lightcurves for typical novae (left) and the specific Nova Vulpeculae 1968 after \citet{Isles1972} (right) with our highlighting of the first ten days: the slight variability ($4.9^{+0.2}_{-0.3}$~mag) would not be recognized by naked eye observers. The $t_3$ time would still be $\sim100$~days for a fast nova ($N_a$-type).}
	\label{fig:lightcurve}
\end{figure*}
The Qintianjian report explicitly states that it remained "calm" for (at least) ten days. We interpret this description as referring to brightness. 

This indicates that its visible duration might be much longer than the ten days recorded. 

After ten days of observations of Qintianjian, the officials with no expertise in astronomy were still able to easily recognize the star, indicating that its brightness was still brighter than 2~mag (to recognize the color) and probably even of negative magnitude to fit the description “calm”. The position given in the text is in the middle between the bright stars Vega, Deneb and Altair so that Vega’s brightness (0~mag) was visually comparable. Novae typically start to decline quickly after their appearance. Assuming that the initial brightness was (at least) $-4$ mag, the interpretation of negative magnitudes for (at least) ten days suggests that the time $t_3$ of decline by 3~mag was longer than 10 days (as the brightnesses of Venus and Saturn can be clearly distinguished and such a difference would unlikely be described as 'calm'). Very fast novae are, thus, almost excluded (but not all N$_a$-type novae, cf. Fig.~\ref{fig:lightcurve}), but all other types of stellar transients are possible.  

\subsubsection{Position and Search Field}
The position is described relative to the constellation \textit{Niandao} and within the lunar lodge \textit{Dou}. The lunar lodge (or mansion) is the section of right ascensions from $\phi$~Sgr (272$^\circ$RA$_{1408}$, today 305$^\circ$RA) to $\beta$~Cap (297$^\circ$RA$_{1408}$, today 281$^\circ$RA). The detailed information on a lunar lodge and the word “measurement” suggests that the observers at the time used instruments to observe the star. According to \citet{chu2022}, the Qintianjian in early Ming consisted of many astronomers and a huge variety of sophisticated instruments, with an average observational error of about 5 arcminutes. With this accuracy of the position measurement, the statement of a non-moving object can be taken seriously, even in modern terms. 

\begin{figure*}[!h]
	\centering
	\includegraphics[width=0.85\linewidth]{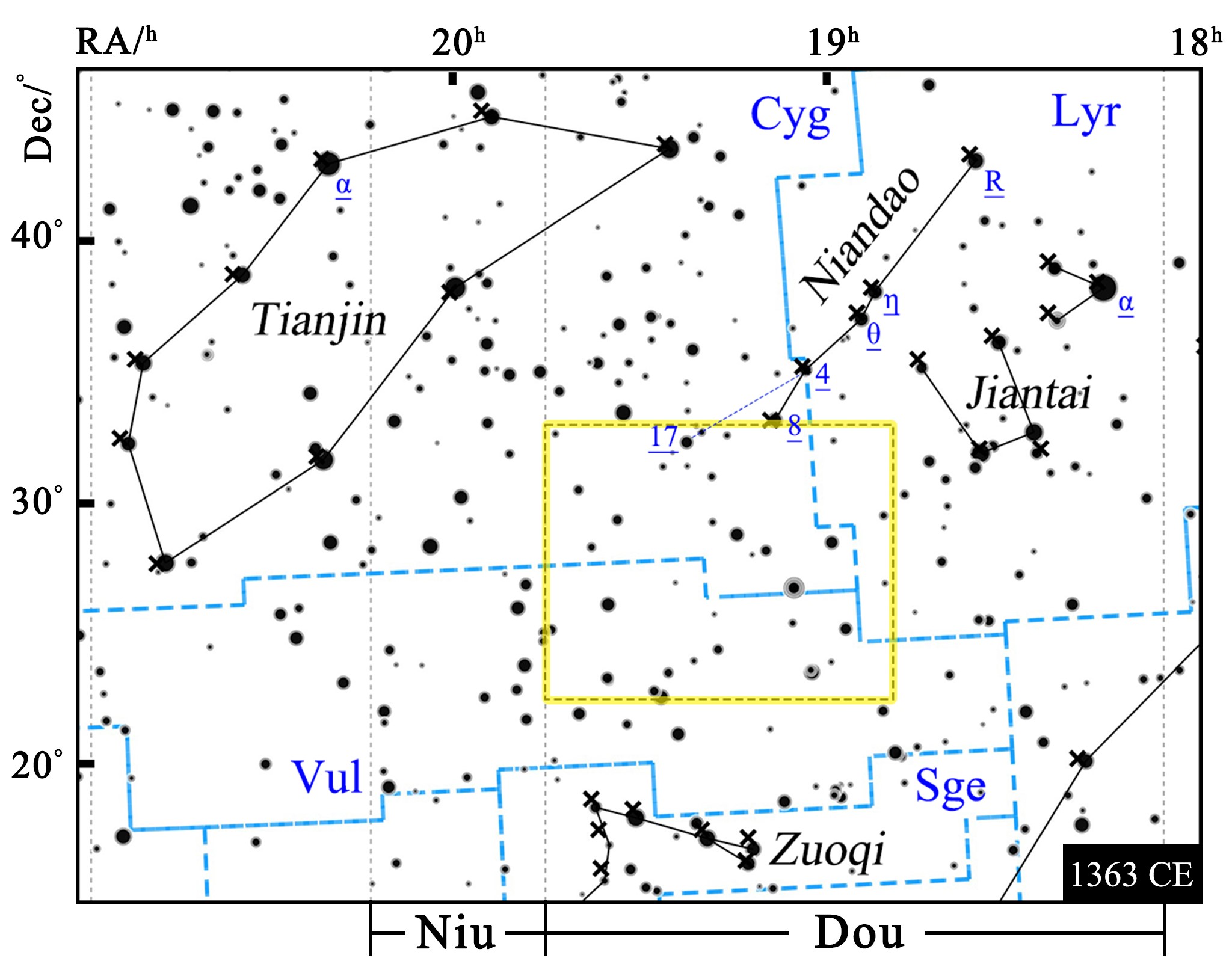}
	\caption{Search Field of the 1408 (Super)Nova. The vertical dashed lines represent the boundaries between different lodges. We used coordinates (RA/DEC)$_{1363}$ as we want to show the constellations in their view of this epoch. The modern coordinates of our field (RA/DEC)$_{2000}$ are given in Tab.~\ref{tab1}.}
	\label{fig:fig1}
\end{figure*}

The asterism of \textit{Niandao} in the Ming Dynasty is considered to consist of the stars 13~Lyr, $\eta$~Lyr, $\vartheta$~Lyr, 4~Cyg and 8~Cyg. Previous researchers used the Qing dynasty star map as a reference, identifying 17 Cyg as the “Southern Star of the Niandao”. However, based on the star catalog observed at around 1363 CE (calculated by \citet[75--80, 220]{yang2023}), it is clear that the Southern Star should be the 4.7~mag star 8~Cyg (HIP 96052) (see Fig. \ref{fig:fig1}, with observational data marked by crosses). Southeast of it is the Milky Way. 

Fig.\ref{fig:fig1} displays the historical and modern map, and the derived search field: It lies in the middle of a huge, constellation-free space of 115 square degrees, bordered by the constellations \textit{Tianjin} and \textit{Zuoqi}, cut by the boundary of the lodge \textit{Dou}. The remaining space is still 50 square degrees wide and in the Milky Way.

\textbf{Remark:} Both the \textit{Ming Taizong shilu} and the \textit{Ming shi} record the guest star's position as "southeast of the \textit{Niandao}." Still, the Memorial of Congratulations suggests it was located south of the \textit{Niandao}. Therefore, the actual position may be slightly to the south-southeast of \textit{Niandao}. 

To be cautious, we will search the southeastern and southern regions of the \textit{Niandao} within the boundaries of the lodge \textit{Dou} (as indicated by the dashed rectangle in the figure). The upper boundary corresponds to the declination of 8 Cyg in 1408, while the left edge of the lodge \textit{dou} marks the left (east) boundary. The right and bottom boundaries are slightly closer to the constellations \textit{Zuoqi} and \textit{Jiantai} to allow for a sufficient margin of error. This area encompasses approximately 100 square degrees within the Milky Way, and it is anticipated that many objects will be found within this region. Thus, we expect to find many objects in this area.   

\subsection{Astrophysics of the area} \label{Astrophysics of the area}
We performed a query in SIMBAD using the coordinates given in Table \ref{tab1} and obtained an expectedly large number of objects of all types. In total, there are 32 SNRs in both fields, 53 potentially interesting stars (CVs, X-ray binaries, WR-stars), and 146 nebulae classified as “planetary nebulae” or “PN candidates.” The latter is a “default category” for unknown objects; some might be SNRs or nova remnants, which needs to be investigated by spectroscopy and the measurement of expansion rates. As a popular example, the SNR of Kepler's SN had been classified as a potential “planetary nebula” prior to \citet{Frew2013}. Spectroscopy easily distinguishes SNRs, but only tediously, the spectra of planetary nebulae can be distinguished from that of a nova remnant as they show the same forbidden line \citep{Kwok2007}. The line ratios of the forbidden oxygen, nitrogen and sulfor lines [O\,III], [N\,II], and [S\,II] to hydrogen lines are not always convincing ([O III] 5007 / H$\beta$ should be larger for PNe ($>3-10$) than ($<1-3$) for nova remnants, while in contrast, [N II] 6583 / H$\alpha$ should be much smaller $0.3-1.5$ for PNe versus $>1$ or even $>3$ for novae; [S II] 6716+6731 / H$\alpha$ is typically small $<0.4$ for PNe and large $>0.5-1$ for novae). However, the ranges are not always sharp, which sometimes led to misidentifications \citep{Miszalski2016,Shara2017b,Goettgens2019}, cf. \citet{Frew2013}. A stronger distinction criterion is the much higher speed of gas in nova remnants ($20-40$~km/s, as opposed to $100-1000$~km/s and more), which can be determined with the line widths (FWHM). Another option would be the distinction by expansion rates (differing by factors of 10 to 30) due to the different initial momentum of thermonuclear runaway eruptions and red giant pulsation-driven ejecta \citep{HoffmannVogt2020a}. Yet, some exceptionally fast PNe and the slowest nova shells could also be mixed up, which may contribute to the number of misclassifications. 

Real planetary nebulae are irrelevant to our research interest, except for some rare phaenomena of so-called rebirth \citep{Bloecker2003,Herwig2006,2011ApJ...743L..33M,Todt2015} with much longer durations than observed here, or diffusion induced novae \citep{MillerBertolami2011}. Therefore, we need to check the catalog for misclassified objects.  

\begin{table}[ht]
\centering
\caption{Search field and (SIMBAD) search results of possible objects}  
\label{tab1} 
\begin{tabular}{|ccc|cc|cc|}
\hline
 && & \multicolumn{2}{c|}{\textbf{south}} & \multicolumn{2}{c|}{\textbf{southeast}} \\
\cline{4-7}
&& & max & min & max & min \\
\hline
 search & coordinates: &RA$_{2000}/ ^\circ$   & 305.3 & 289   & 302  & 292  \\
 & &DEC$_{2000}/ ^\circ$   & 34.5  & 21.5  & 39   & 24   \\
\hline
  results by &object types: & SNRs  & \multicolumn{2}{c|}{32} & \multicolumn{2}{c|}{21} \\  
 &&CVs   & \multicolumn{2}{c|}{27} & \multicolumn{2}{c|}{33} \\  
 &&XBs   & \multicolumn{2}{c|}{3}  & \multicolumn{2}{c|}{3}  \\
 &&WR    & \multicolumn{2}{c|}{6}  & \multicolumn{2}{c|}{7}  \\
 &&PN?   & \multicolumn{2}{c|}{62} & \multicolumn{2}{c|}{36} \\
 &&PN    & \multicolumn{2}{c|}{119} & \multicolumn{2}{c|}{84} \\
\hline
\end{tabular}
\end{table}

Some planetary nebulae and so-called `PN candidates' are also found in this area in the Milky Way:  PN Ra 13, TYC 2674-1567-1, 2MASS J19595642+3048238, G068.2+00.9, G066.4-00-0, HIILBN 070.13+01.71, PN G070.4+00.7, G064.9+00.7, G064.1+00.7, G062.2+01.1, G061.8+00.8, G062.1+00.1, G065.8+05.1 are rather certainly not nova remnants. The nebulae G070.9+02.2 and RAS 20032+3212 are close to the described position but in the wrong lunar mansion, so we exclude them.

\subsubsection{Are there supernova candidates? }
Most of the supernova remnants are undetermined in age or, with age estimates of some $10^4$ years, yet too old. Three of them have a pulsar in or near their center. Two cases may be interesting to study further regarding the event in 1408 (roughly 600 years ago). Both of them are too old at first glance, but may perhaps get into the age range if the age estimate is changed.  

\begin{itemize}
  \item The SNR G063.7+01.1 (filled center SNR, \textgreater8000 years, PWN but no PSR known: perhaps 3XMM J194753.4+274357 (CXO J194753.3+274351) driving the PWN; ManitobaCat).
  
 \item The SNR G067.7+01.8 is even more interesting because it is much younger, differing from the event only by a factor of two, as the youngest possible age is 1500~yr. Still, the upper age limit of 13,000~yr also excludes it as a candidate. There is no PSR known in this nebula with a radio-bright center. 
\end{itemize}

\subsubsection{Cataclysmic and other contact binaries}
Additionally, there are roughly 20 bright CVs, plus eight bright symbiotic and X-ray binaries. The list of the brightest and most interesting objects in the search field is given in Tab.~\ref{tab:CVs}.

\begin{table}[h]
{\centering
\label{tab:CVs} 
\caption{Nova candidates within the search field: VSX data and Gaia DR3 parallaxes \citep{GaiaDR3}.}
\includegraphics[width=\linewidth]{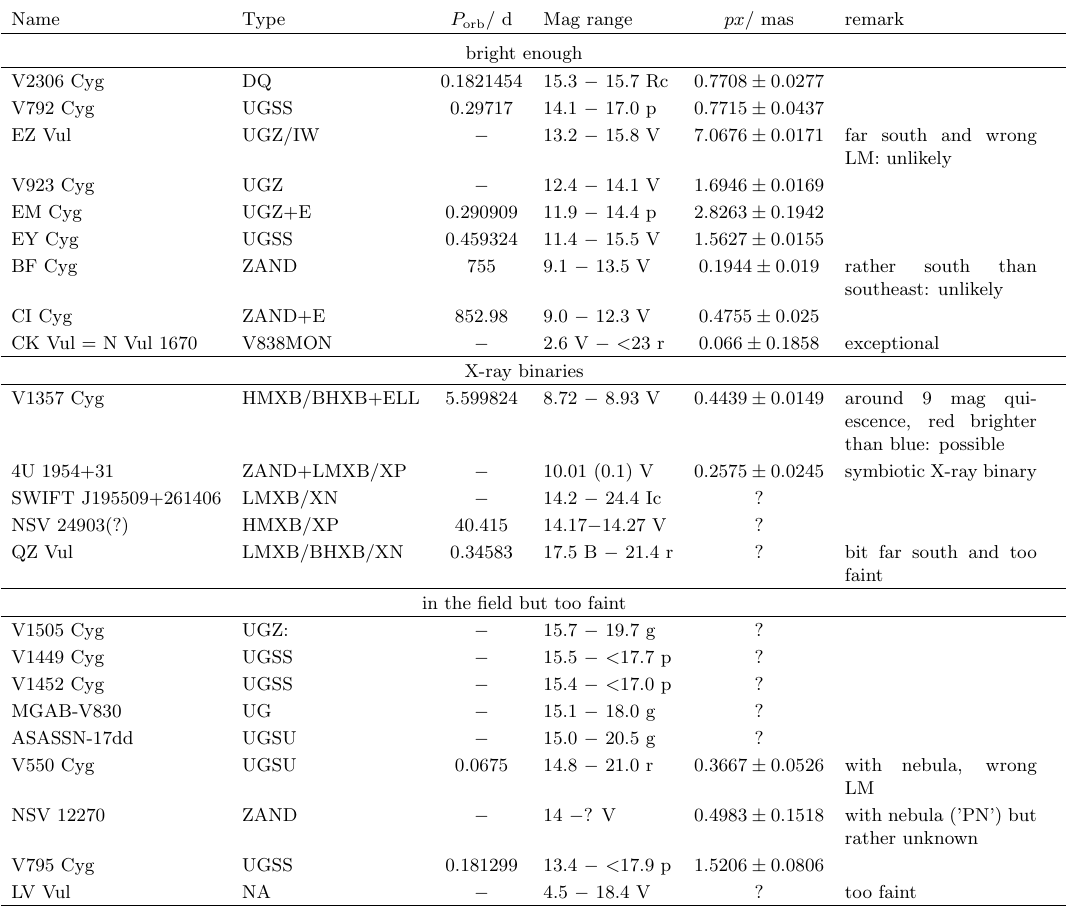} \\}
\end{table}

\subsubsection{X-ray binaries}
 No case is known that a low-mass or high-mass X-ray binary has ever erupted as a nova. Yet, X-ray binaries are possible sources of eruptions, e.\,g. tidal disruption events (TDEs) or other high-energy emission (that may or may not also flare up the system in visible light). 

X-ray binaries are contact binaries whose compact companion might be a white dwarf (like in novae), a neutron star, or a black hole. Thus, the X-ray emission may originate from a variety of different physical effects. Likewise, any sort of eruption from these systems may have different causes. 

Tidal Disruption Events (TDEs) were first observed in high-energy astrophysics (ROSAT data), but recently, optical surveys like ZTF and ASAS-SN have also detected several of them \citep{Yao2023}. Typical known cases still originate from systems with black holes, even supermassive black holes in the centers of galaxies \citep{Gezari2021, Yao2023}. Yet, occasionally stellar binaries with a black hole companion are also discussed, although frequently rejected \citep{Rivinius2020,Safarzadeh2020,Bodensteiner2020}. As light curves of TDEs show a characteristic (observable) incline of brightness (in contrast to novae and supernovae where the process of flaring up is typically received as immediate), we consider them less likely options for historical observations, yet, not automatically excludable. The duration of the visibility of TDEs in the known systems ranges from months to years, and light curves in \citet[fig.\,4]{Gezari2021} and \citet[fig.\,8]{Yao2023} suggest stability over the first ten days after peak. Still, the long phase of rising to peak brightness would have been noticed and probably described as appearance or growing of the object.

 For completeness, we mention the X-ray binaries, although we consider such a scenario rather unlikely for our historical observation. 

\subsubsection{Nova candidates}
A few known novae are in this Milky Way area. The known novae QV Vul (Nova Vul 1987) and NQ Vul (Nova Vul 1976) are slightly too far west, and with historical eruption peaks fainter than 6~mag, they are also not bright enough to produce a phenomenon as described in the Ming Dynasty. QU Vul (Nova Vul 1984b) could match the position, but with 5.3~mag in peak, this N$_a$-type nova would also be too faint \citep[260]{HoffmannVogt2021}. The only potential candidate would be LV Vul, the Nova Vulpeculae 1968, see Fig.~\ref{fig:lightcurve} \citep{Isles1972}. The light curve of this fast nova (N$_a$-type) appeared stable for the first ten days, and \citet{downes2000} and \citet{Hachisu_2019} give the time of decline by 3~mag $t\textsubscript{3} \sim43$ days. However, the peak brightness was estimated roughly 4.5~mag in V. Based on the known variability of peak brightness of recurrent novae ($\pm2$ to $4$~mag, see \citet[fig. 3]{HoffmannVogt2022}) we would not expect it to achieve negative magnitudes. So, this object is highly unlikely to cause the 1408 observation. Still, novae with recurrence periods of more than a century (in this case, it would be 1968$-$1408 $=$ 560 years) are unknown \citep{Darnley2021,HoffmannVogt2022}, and the behavior of objects like this may differ from the known cases. We denote this for future research.

With typical amplitudes of 11 to 13 mag and maximum amplitudes of 16 mag for classical novae \citep{vogt2019possibilities,Rosenbush2020}, smaller amplitudes for symbiotic novae \citep{Munari2019}, and the given stability of the visual appearance in the first ten days, here we consider identifications either with a Z And-type star, a dwarf nova (UG-type), or a DQ Her-type intermediate polar as candidates. DQ~Her-type systems permit novae whose light curve is rather stable at the beginning \citep[fig.\,2]{Strope2010a}, but the stability might last even longer than 10~days. 

Z And-type stars are symbiotic. They could be a combination of a red giant and a white dwarf, like all slow (N$_c$-type) novae and some observed recurrent novae. Their orbital periods are also similar to those of N$_c$-type novae. Cataclysmic variables (CV) of dwarf nova-type (UG) could permit a classical nova, and DQ Her-type contact binaries have light curves with a plateau at the beginning \citep{Strope2010a}. In total, we found nine candidates within these main groups (see Tab.~\ref{tab:CVs}, uppermost section). In the case of the CVs, the range of their orbital periods coincides with the range of those of known N$_b$-type novae. 

With typical amplitudes \citet[fig. 1]{vogt2019possibilities}, the quiescence of a CV should be $8\pm4$ to reach $-4\pm3$~mag in eruption. Assuming maximum amplitude (16~mag), the progenitor system should have at least 13~mag in quiescence. None of the star in Tab.~\ref{tab:CVs} qualifies for a typical case, and the extreme amplitude yields only four candidates: the dwarf novae EM Cyg and V923 Cyg, and the symbiotic systems BF Cyg and CI Cyg. As novae from symbiotic systems have smaller amplitudes, none of these is likely.

Furthermore, assuming typical absolute magnitudes of novae, most objects in Tab.~\ref{tab:CVs} would be ruled out: \citet{1981PASP...93..165D} assumed $M_V=-6\dots-7$~mag for slow novae and  $M_V=-8\dots-11$~mag for fast ones. More recent studies seem to calculate with a value of $M_V=-8.5$~mag \citep{Gilmozzi2003,kok2010,Rosenbush2020,2024ApJS..272...28C}. It has even been proposed to use classical novae as standard candles \citep{Gilmozzi2003} as their absolute magnitudes in peak, and even 15~days after peak, seem to be always the same, as the same physical processes cause them. The very detailed study by \citet{kantharia2017} concludes for novae in our Galaxy, M31 and M87 $M_V=(-8\pm2)$~mag (or $-6$ to $-10$), again, roughly the range given by \citet{1981PASP...93..165D}. The distance modulus would then lead to an estimated distance of the classical nova between 63 and 630~parsecs, and a parallax between 1.6 and 160~mas. The Gaia parallaxes displayed in Tab.~\ref{tab:CVs} would then only allow EM~Cyg, V923 Cyg and EZ~Vul as candidates (plus the nine objects for which no Gaia data is available). 

\subsubsection{The peculiar object: CK~Vul}
\begin{figure*}[h]
	\centering
	\includegraphics[width=0.88\linewidth]{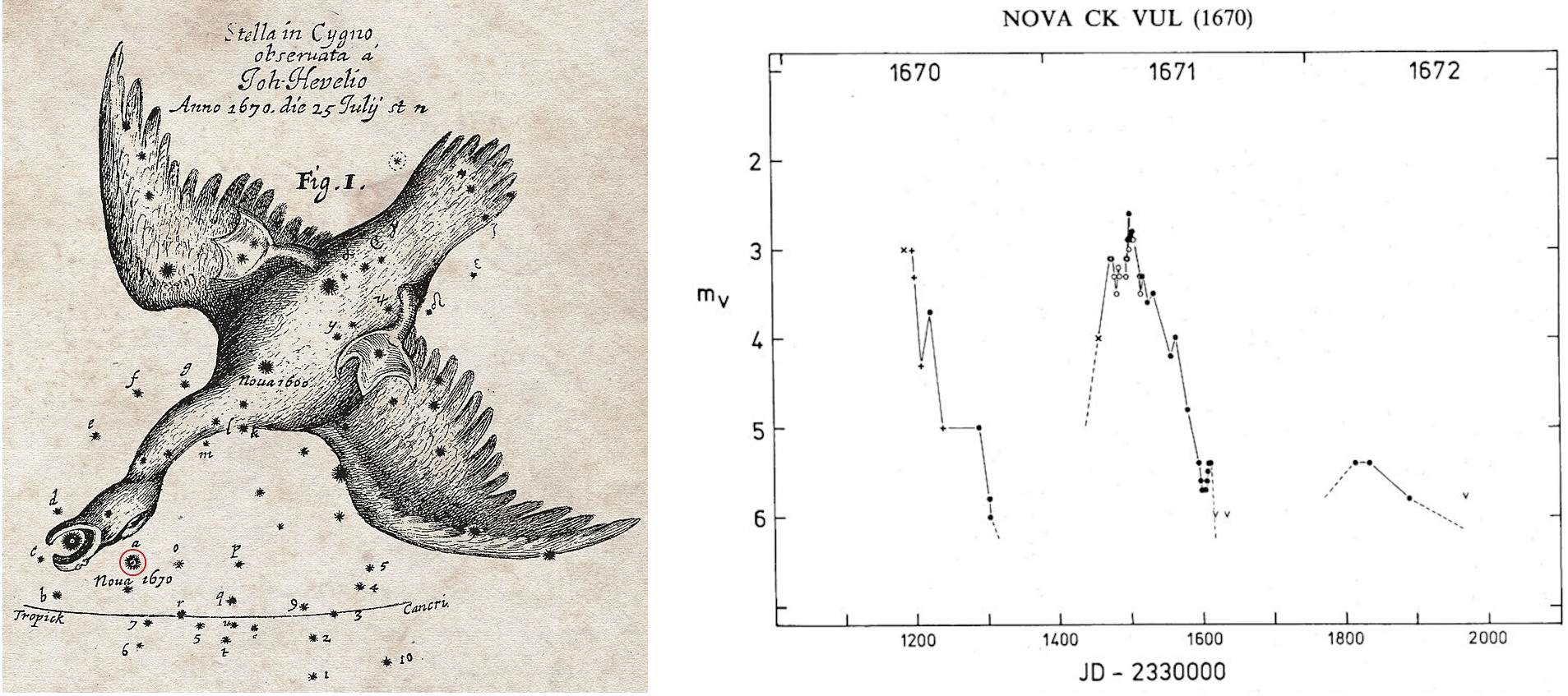}
	\caption{Observation drawn by \citet{hevel1670} (left) and light curve (right) of the Nova Vul 1670--1672 by \citep{shara1985} from observations of Anthelme, Cassini, Hevelius and various anonymous members of the Académie des Sciences, Paris.}
	\label{fig:hevelNova}
\end{figure*}

Despite all of the above facts that may rule it out, we consider the special case CK Vul worth some consideration: it has been suggested as a nova by \citet{shara1982recovery}, usually considered as counterpart of Nova Vulpeculae 1670 observed by \citet{hevel1670} and his contemporaries. This event exhibited rather unusual behavior, as it was reported with three separated maxima in the years 1670, 1671 and 1672 (see the light curve reconstructed by \citet{shara1985}, our Fig.~\ref{fig:hevelNova}, also \citet{shara1982recovery,hajduk2007}.

 \citet{MillerBertolami2011} suggested that the simulated scenario for a `diffusion induced nova' (DIN) offers an explanations for CK Vul, and explains the observed properties of [WN/WC]-central stars of planetary nebulae. They seem imply that the event in 1670--'72 was a rebirth scenario of a CSPN, already suggested by \citet{Evans2002}. However, \citet{Hajduk2013} with optical observations, and \citet{Evans2016}, based on new Spitzer infrared observations, reject the possibility that CK~Vul is a remnant of any type of nova: neither could it be a classical nova remnant in hibernation, a very late thermal pulse, a Luminous Red Variable, such as V838 Mon, nor a diffusion-induced nova. In their view `The true nature of CK Vul remains a mystery'. The same conclusion is reached by \citet{Banerjee2020}, who derive an absolute magnitude of $M_v=-12.4^{+1.3}_{-2.4}$ of the event in 1670 and a distance of $3.2^{+0.9}_{-0.6}$~kpc for CK~Vul (which is not compatible with Gaia DR3 data). They speculate if CK~Vul forms part of a peculiar group of intermediate-luminosity optical transients (ILOTs) that bridge the luminosity gap between novae and supernovae. Summarizing, infrared data does not support \textit{any} nova hypothesis for CK~Vul; the object and event remain peculiar.

 The explanation as a “luminous red nova” (LRN) or V838MON, a stellar collision in a contact binary had been proposed by \citet{kato2003,kaminski2015post} and supported by \citet{Hajduk2013} who stressed that one of the components must have been a white dwarf. More recent research \citep{Tylenda2019,Kaminski2023,Tylenda2024} considers the novae 1670--'72 as a merger of a red giant with a helium white dwarf. Yet, this hypothesis appears brave as helium white dwarfs are extraordinarily rare; only two are known \citep{liebert2004,k+k2010,brown2013}, and they typically are not created in binary systems. 
 
 In their articles, they do not give an accurate date for the merging during the sequence of three maxima in three years. Yet, it is known that the duration of the visibility of all nova and supernova eruptions depends on the surrounding nebulae of gas and dust that reflect the light. The red giant or supergiant in symbiotic binaries like CK Vul always produces an embedding of gas and dust clouds (see prototype of V838Mon-type stars). Today, the stellar remnant is not directly visible because of opaque molecule clouds, and it can be assumed that the emission of the nebula from the symbiotic system had been ongoing for a long time when the merger happened in the 1670s. It is well possible that parts of one or more of the peaks in the 17th century originate from reflections of the energy released by the merger in some nearby cloud filaments. 
 
 Before merging, the binary would have orbited closely, perhaps permitting a transfer of hydrogen-rich gas from the giant’s atmosphere towards the white dwarf for a long time. So, it is not excluded that about 200 years prior to the merging event, a classical nova outburst had occurred in CK Vul, observed by Chinese astronomers \citep{HoffmannVogt2021}. In DQ~Her-type intermediate polars (which we preferred as candidates due to their broad peak \citep{Strope2010a}), the magnetic field disrupts the accretion disc and a dust belt, which may align with CK Vul's bipolar nebula structure \citep{Hajduk2013,Banerjee2020}. A classical nova of the system may, thus, show similar behavior and match the Chinese observation in 1408. 
 
 During this hypothetical nova eruption, most of the outer hydrogen layer of the white dwarf (accumulated during previous $\sim$100,000 years) was transformed into helium, this way causing the combination between a giant and a helium white dwarf that finally merged according to \citet{Tylenda2019}. Thus, our hypothetical scenario would explain the presence of a (rare) helium white dwarf in the CK~Vul system. \citet{yaron2005} estimates the time between outbursts within the same system based on mass transfer rates onto a white dwarf of 0.4~$M_\odot$ to the order of $10^6$~yr, which would exclude a recurrence after only $\sim250$~years. However, in our scenario, the second event would be of a different nature and different physics than the first one. The first event would be a thermonuclear runaway on the surface of an ordinary white dwarf, turning it into a helium white dwarf, and the second one a merger of this with its contact companion.

 Although the spacial distance of CK Vul derived from most recent Gaia data seems to contradict its potential to become visible as a classical nova of typical absolute magnitudes, the preliminary idea of a nova of CK Vul in 1408 should be investigated,as everything about it seems to be exotic: the merger, the helium white dwarf and the facts that (a) the nova prior to the merger would have been brighter than the merger, and that (b) only eight members of the V838MON class are registered in the VSX, three of them uncertain. Given that five stars of this type are distributed over the entire sky (41,253$^\circ$²), an accidental coincidence of CK Vul with a historical observation within a field of only $\sim100\degr^2$ is rather improbable (0.2\,\%). Furthermore, our new finding of an observational record from 1408 confirms the hypothesis of a stellar transient in the area. 

\section{Conclusion} \label{sec:conclusion}
The 1408 CE "guest star," documented in Chinese historical records, is re-evaluated as a probable classical nova based on critical analysis of original Ming Dynasty sources and modern astrophysical data. Key features of the observation (stationary position over ten days, stable light curve and luminous appearance) align with characteristics of stellar transients rather than cometary phenomena. Positional constraints localize the event in the Cygnus-Vulpecula region, where we identify CK Vul, an LRN remnant from 1670–-1672, as a possible candidate. The symbiotic progenitor system could have permitted a classical nova eruption in 1408, preceding its later merger event \citep{kato2003,kaminski2015post}. While other supernovae, cataclysmic variables and planetary nebulae within the search field were examined, most lack the required age, brightness, or light-curve stability to match the historical account. Further observations of all discussed objects (EM~Cyg, V923~Cyg, EZ Vul, and CK Vul) are needed to verify the hypotheses proposed in this paper. If these stars are not the counterpart of the 1408 transient event, then the counterpart may still remain undiscovered.

This study highlights the synergy between meticulous historical scholarship and modern astrophysical techniques in resolving ambiguities surrounding ancient transients. The 1408 event stands as one of the earliest well-documented nova candidates, offering a rare opportunity to probe pre-telescopic stellar phenomena and their modern counterparts. Future high-resolution observations of CK Vul’s remnant and spectroscopic studies of its circumstellar environment could test the hypothesized connection to the 1408 eruption. Such interdisciplinary efforts underscore the enduring value of historical records in constructing a comprehensive timeline of Galactic transient activity.  

\section*{Authors' Contributions}
Yang and Hoffmann performed the research together. Hoffmann initiated the research due to open questions from earlier projects. Yang discovered the 1408 observational report and analyzed the color, brightness, position, and related information. Hoffmann had searched for possible (super)nova counterparts of the event in 1408 in earlier publications. Thus, when she designed this paper, she consulted her former co-author, the expert on cataclysmic variables and novae, Nikolaus Vogt for a closer discussion of the candidates.

\begin{acknowledgments}
\section*{Acknowledgements}
We thank Prof. Yunli Shi for supporting two guest stays of SMH in Hefei in 2024, and the financial aid from the National Social Science Fund of China (16ZDA143), the Austrian Academy of Science (Otto Neugebauer Fund), and the China Postdoctoral Science Foundation fellowship (2024M753101).

Thankfully, we made use of the stellar databases SIMBAD \citep{Wenger2000} and the Variable Star Index (VSX) of the AAVSO \citep{Watson2006}, and the supernova remnant catalog at the University of Manitoba \citep{Ferrand2012}. 

Furthermore, we thank our great reviewer for their thoughtful and very helpful suggestions to improve the readability of our contribution. 
\end{acknowledgments}


\bibliography{1408}{}
\bibliographystyle{aasjournal}

\end{document}